\documentstyle[12pt]{article}
\begin{document}

\noindent
{\Large \bf White Dwarfs and Dark Matter}

\smallskip

\noindent
{\large Accepted for publication as a \it Technical Comment \rm 
in \it Science\rm, 
in response to Oppenheimer et~al. (2001,\it 
Science\rm, 292, 698).}

\medskip
\hfil\bf B.~K. Gibson \rm\break

\vspace{-5.0mm}
\hfil Centre for Astrophysics \& Supercomputing\break

\vspace{-5.0mm}
\hfil Swinburne University\break

\vspace{-5.0mm}
\hfil Mail \#31, P.O. Box 218\break

\vspace{-5.0mm}
\hfil Hawthorn, Victoria, 3122, Australia\break

\hfil\bf C. Flynn \rm\break

\vspace{-5.0mm}
\hfil Tuorla Observatory\break

\vspace{-5.0mm}
\hfil Turku University\break

\vspace{-5.0mm}
\hfil V\"ais\"al\"antie 20\break

\vspace{-5.0mm}
\hfil FIN-21500, Piikki\"o, Finland\break
\bigskip

Based upon the identification of 38 white dwarfs with halo kinematics, in a
survey covering 10\% of the sky near the south Galactic pole, Oppenheimer {\it
et~al.}  ({\it 1~}) argue that at least 2\% of the dark matter in the Galaxy
has now been detected directly.  
Put into context, the Oppenheimer {\it et~al.}
result implies that the {\it stellar remnant} mass of the halo may be
comparable to that of the {\it entire} disk of our Galaxy.  
If true,
this finding has crucial consequences for understanding the formation and
evolution of the Milky Way.
Careful examination of the Oppenheimer {\it et~al.} results leads us to 
conclude that they have overestimated the density of white dwarfs with halo
kinematics.

Oppenheimer {\it et~al.} derive their local white dwarf density $n$ 
via the $1/V_{\rm max}$ technique ({\it 2~}). The equation that applies 
for a survey covering 10\% of the sky is
$$
n = \sum_{i=1}^{38}V_{i,\rm max}^{\,-1} 
\approx 2.4\,d_{i,\rm max}^{\,-3}\;\;{\rm pc}^{-3}
$$
where $V_{\rm max}$ represents
the maximum volume in which the survey could have found each of the 38 white
dwarfs listed in Table~1 of ({\it 1~}) and
$d_{\rm max}$ is the distance in parsecs which determines $V_{\rm max}$.
Oppenheimer {\it et~al.} considered two
relations for $d_{\rm max}$, one depending upon
the limiting magnitude of the survey
R59F$_{\rm lim}$ and the luminosities M$_{i,\rm R59F}$ of each of the 38 white
dwarfs, and one depending upon the inferred distance $d$ and observed proper
motion $\mu$.

Using equation~1 and the 38 white dwarfs in their sample,
Oppenheimer {\it et~al.} derived a white dwarf number density 
$n$=1.8$\times$10$^{-4}$\,pc$^{-3}$.  We rederived $n$, employing
the above relation, the data tabulated in their Table~1, and the 
identical $d_{i,\rm max}$ criteria used in their analysis, and
found $n$=1.54$\times$10$^{-4}$\,pc$^{-3}$.  Moreover,
Oppenheimer {\it et~al.} assumed a typical white dwarf mass of
0.6\,M$_\odot$, which in combination with their derived number density, 
resulted in a local mass density of
1.1$\times$10$^{-4}$\,pc$^{-3}$\,M$_\odot$\,pc$^{-3}$.  
By contrast, in metal-poor systems
such as globular clusters - which would be expected to mimic to some degree 
the patterns in the 
Galactic halo proper - the typical white dwarf mass is 
0.51$\pm$0.03\,M$_\odot$
({\it 3~}).  That average mass, combined
with our recalculated number density, results
in a local white dwarf mass density of
0.79$\times$10$^{-4}$\,M$_\odot$\,pc$^{-3}$, 30\% below that found
by Oppenheimer {\it et~al.}  Even that revised density should be viewed with
caution, because 32\% of the density inferred
from our reanalysis is being driven by only 8\% of the sample - i.e., 
three white dwarf candidates (LP651-74, WD0351-564, and WD0100-567)
contribute 19\%, 7\%, and 6\% to the total, respectively.

Oppenheimer {\it et~al.} derived a mean V/V$_{\rm max}$ of 0.46,
assuming a limiting apparent magnitude of R59F$_{\rm lim} = 19.80$,
and suggest that a more appropriate R59F$_{\rm lim}$ should then be
19.70, in order to yield $<$V/V$_{\rm max}$$>$=0.50 (expected for 
a uniform distribution).  Using R59F$_{\rm lim} = 19.70$ and
$<$m$_{\rm WD}$$>$=0.6\.M$_\odot$, they arrived at their quoted
result of
1.3$\times$10$^{-4}$\,pc$^{-3}$\,M$_\odot$\,pc$^{-3}$.
Our analysis, by contrast, leads to a {\it mean} V/V$_{\rm max}$ of 0.44.
At face value, that result would imply that R59F$_{\rm lim}$ should be
adjusted to
19.55, in order to recover $<$V/V$_{\rm max}$$>$=0.50.  Such an adjustment,
however, has little effect,
increasing the inferred local white dwarf mass density
from 0.79$\times$10$^{-4}$\,M$_\odot$\,pc$^{-3}$ to
0.88$\times$10$^{-4}$\,M$_\odot$\,pc$^{-3}$.

Indeed, it should be stressed though that in {\it both}
of the above cases, although the mean V/V$_{\rm max}$ was below 0.50 for
R59F$_{\rm lim} = 19.80$, the {\it median}
V/V$_{\rm max}$ was exactly 0.50 - that is, there is little reason to
modify R59F$_{\rm lim}$ from 19.80, to either
19.70 or 19.55.  More important, perhaps, the
increase in the {\it mean} R59F$_{\rm lim}$ comes about by increasing
V/V$_{\rm max}$ for several of the white dwarfs to values in excess of
unity, a physical impossibility.  The problem lies in the non-normal
distribution 
of V/V$_{\rm max}$ for the sample, in which 13 of the 38 white 
dwarfs have V/V$_{\rm max}$$<$0.2.  Modifying R59F$_{\rm lim}$ has little
impact upon those 13 white dwarfs, but does increase V/V$_{\rm max}$ for
those white dwarfs whose ratios are larger.  
We conclude that modifying R59F$_{\rm lim}$ from 19.80 to 19.55, or
19.70, is misleading, and leads to unphysical values for V/V$_{\rm max}$,
and thus
favour our result for the local white dwarf
mass density of 0.79$\times$10$^{-4}$\,M$_\odot$\,pc$^{-3}$, 40\% below the
1.3$\times$10$^{-4}$\,M$_\odot$\,pc$^{-3}$ value found by
Oppenheimer {\it et~al.} ({\it 1~}).

Instead of a factor of ten excess relative to the mass density 
expected from a standard initial mass
function ({\it 4}), our revision puts the excess at a (still significant)
factor of six.
Oppenheimer {\it et~al.} quoted a white dwarf halo mass fraction 
of 2\% (rounded up from 1.6\%), whereas our results imply a fraction of
1.0\%.  Both of these fractions, however,
assume a local dynamical halo mass density of
8.0$\times$10$^{-3}$\,M$_\odot$\,pc$^{-3}$, based upon the
Oppenheimer {\it et~al.} reading of the work of Gates {\it et~al.} ({\it
5~}).  Our reading of Gates {\it et~al.}, by contrast,
suggests that this local density
should actually be (14$\pm$5)$\times$10$^{-3}$\,M$_\odot$\,pc$^{-3}$, a
normalisation that effectively reduces the local white dwarf halo mass
fraction from 1.0\% to 0.6\% (and that 
would change the Oppenheimer {\it et~al.}
result from 1.6\% to 1.0\%).

Finally, the local 
density of halo white dwarfs claimed by Oppenheimer {\it et~al.}
does not seem consistent
with the combined results of deep proper motions surveys taken to date ({\it
6~}). These surveys utilise the so-called {\it reduced proper motion} to find
dark halo objects, since this very clearly separates them from objects of the
disk and stellar halo. The absolutely faintest white dwarf in the Oppenheimer
{\it et~al.}  sample is at M$_{\rm R59F}$=15.9 (WD0351-564). If we 
conservatively allow
2\% of the dark halo to be in the form of white dwarfs of this luminosity, then
we would expect very significant numbers of them to have been found in existing
surveys. In the Luyten Half Second survey (LHS), with a limiting R-band
magnitude of 18.5 and a proper motion window of $ 0.5 < \mu < 2.5$ arcsec/yr,
we would expect some 15 dark halo white dwarfs with reduced proper motions in
the range 23.5 to 25.5, 
whereas only a few such objects are known. The Oppenheimer
et al survey covers an almost identical volume for dark halo objects as the LHS
(which is less deep but covers about half the sky). We would also expect some
15 objects like WD0351-564 in the Oppenheimer {\it et al.} survey (adopting
their apparent magnitude limit and window for detectible proper motions) and in
the same reduced proper motion range, whereas only 3 such objects were
found (F351-50, WD0351-564, LHS542). 
This strongly suggests that the local density of these objects has been
overestimated. We estimate a 2$\sigma$ {\it upper limit} for the contribution
of objects of the luminosity of WD0351-564 of 0.5\% to the local dark halo
density based on a conservative estimate that not more than 5 such objects are
seen in the LHS and Oppenheimer {\it et al.} surveys combined.  This 
independent assessment agrees with the 0.6\%-1.0\% derived above.

In sum,
we argue that the objects found by Oppenheimer {\it et al.} have a local
density that is a factor of two to four less
than their claimed detection of 2\% of the Galactic dark matter. 
White dwarfs, at present, represent the most exciting (and least radical) of
candidates for a component of the Galactic dark matter, even though 
the indirect problems associated with this identification remain daunting
({\it 7~}).  The Oppenheimer {\it et
al.} result shows that we may be close to experimental confirmation or
rejection of this proposal.
Deep proper motion surveys for fast-moving
faint objects could well resolve this issue within a few years.

\bigskip
\bigskip
%\newpage

\noindent
{\Large \bf References\rm}

\noindent

\begin{enumerate}

\item B.~R. Oppenheimer, N.~C. Hambly, A.~P. Digby, S.~T. Hodgkin, D.~Saumon,
{\it Science} {\bf 292}, 698 (2001).

\item M.~A. Wood, T.~D. Oswalt, {\it Astrophys. J.}  {\bf 490}, 870 (1998).

\item H.~B. Richer, {\it et~al}, {\it Astrophys. J.} {\bf 484}, 741 (1997).

\item A. Gould, C. Flynn, J.~N. Bahcall, {\it Astrophys. J.}  {\bf 503}, 798
(1998).

\item E.~I. Gates, G. Gyuk, M.~S. Turner, {\it Astrophys. J.}  {\bf 449}, L123
(1995).

\item C. Flynn, J. Sommer-Larsen, B. Fuchs, D. Graff, S. Salim, {\it Mon. Not.
R. Astron. Soc.}, {\bf 322}, 553 (2001). 

\item B.~K. Gibson, J.~R. Mould, {\it Astrophys. J.}  {\bf 482}, 98 (1997).

\end{enumerate}

\end{document}